\documentclass[fleqn,usenatbib]{mnras}

\usepackage{newtxtext,newtxmath}

\usepackage[T1]{fontenc}

\DeclareRobustCommand{\VAN}[3]{#2}
\let\VANthebibliography\thebibliography
\def\thebibliography{\DeclareRobustCommand{\VAN}[3]{##3}\VANthebibliography}

\usepackage{graphicx}	
\usepackage{amsmath}	
\usepackage{amssymb}	

\usepackage{dcolumn}
\usepackage{bm}
\usepackage{amsfonts,slashed,upgreek,color}
\usepackage{multirow}
\usepackage{cancel}
\usepackage{hyperref}

\newcolumntype{C}[1]{>{\centering\arraybackslash}p{#1}}

\newcommand\ee{\end{equation}}
\newcommand\be{\begin{equation}}
\newcommand\eea{\end{eqnarray}}
\newcommand\bea{\begin{eqnarray}}
\newcommand{\bV}{\mathbf{V}}

\newcommand{\bn}{\mathbf{n}}

\newcommand{\bk}{\mathbf{k}}  

\newcommand{\B}{\textrm{B}}
\newcommand{\F}{\textrm{F}}

\newcommand{\HH}{\mathcal{H}}

\newcommand{\I}{I}

\newcommand\Tstrut{\rule{0pt}{2.6ex}}

\newcommand\Bstrut{\rule[-1.2ex]{0pt}{0pt}}

\title[Measuring time distortion]{Measuring the distortion of time with relativistic effects in large-scale structure}

\author[D. Sobral Blanco and C. Bonvin]{
Daniel Sobral Blanco\thanks{daniel.sobralblanco@unige.ch},
Camille Bonvin\thanks{camille.bonvin@unige.ch}
\\
D\'epartment de Physique Th\'eorique and Center for Astroparticle Physics,
Universit\'e de Gen\`eve, Quai E. Ansermet 24, CH-1211 Gen\`eve 4, Switzerland
}

\date{Accepted XXX. Received YYY; in original form ZZZ}

\pubyear{2022}

\begin{document}
\label{firstpage}
\pagerange{\pageref{firstpage}--\pageref{lastpage}}
\maketitle

\begin{abstract}
To test the theory of gravity one needs to test, on one hand, how space and time are distorted by matter and, on the other hand, how matter moves in a distorted space-time. Current observations provide tight constraints on the motion of matter, through the so-called redshift-space distortions, but they only provide a measurement of the sum of the spatial and temporal distortions, via gravitational lensing. In this Letter, we develop a method to measure the time distortion on its own. We show that the coming generation of galaxy surveys, like the Square Kilometer Array, will allow us to measure the distortion of time with an accuracy of 10-30\%. Such a measurement will be essential to test deviations from the $\Lambda$CDM model in a fully model-independent way. In particular, it can be used to compare the spatial and temporal distortions of space-time and to unambiguously distinguish between modifications of gravity and dark fifth forces acting on dark matter.
\end{abstract}




\section{Introduction} 
\label{sec:intro}

Two fundamental puzzles are at the core of current research in cosmology: what is causing the observed accelerated expansion of the Universe; and what is the nature of dark matter. The standard model of cosmology, the $\Lambda$CDM model, postulates that the accelerated expansion is due to a cosmological constant, and that dark matter is a perfect non-relativistic fluid which interacts only gravitationally with the normal baryonic matter. One of the main goal of current and future large-scale structure surveys is to test the validity of the $\Lambda$CDM model, and explore deviations from this minimal scenario.

Deviations from $\Lambda$CDM are often studied separately by different communities: some studies concentrate on models beyond General Relativity (see e.g.~\cite{Koyama:2015vza,Joyce:2016vqv} for a review), assuming that dark matter obey the weak equivalence principle; while other studies explore non-standard dark matter models subjected to a dark fifth force (see e.g.~\cite{PhysRevLett.67.2926,1992ApJ...398..407G,Barros:2018efl,Asghari:2019qld,Archidiacono:2022iuu}), assuming the validity of General Relativity (GR). In practice however, since we do not know if GR is valid and we do not know if dark fifth forces exist, we would like to use observations to distinguish between these different scenarios. More precisely, if we observe deviations from the $\Lambda$CDM model, it is legitimate to ask ourselves if they are due to a modification of the theory of gravity, or to a fifth force acting on dark matter~\footnote{Note that there are also models where gravity is modified and in addition the equivalence principle is broken since dark matter and baryons are coupled differently to the new degree of freedom~\citep{Gleyzes:2015pma,Gleyzes:2015rua}. In other models the equivalence principle is valid at the microscopic level, but the modifications of gravity generate an effective breaking of the equivalence principle due to screening~\citep{Hui:2009kc}.}.

One key observable in large-scale structure surveys is the growth rate of perturbations $f$, that can be measured in a model-independent way from redshift-space distortions (RSD), see e.g.~\cite{eBOSS:2020yzd}. While this growth of structure provides a powerful way of unveiling deviations from $\Lambda$CDM, it cannot unambiguously distinguish between modifications of gravity and dark fifth forces since they impact the way matter is clustering in a very similar way \citep{Song:2010fg,Castello:2022uuu,Bonvin:2022tii}. A key difference between modifications of gravity and dark fifth force models however, is the fact that in modified gravity theories, very generically the two gravitational potentials encoding spatial ($\Phi$) and temporal ($\Psi$) distortions are different~(see e.g.~\cite{Saltas:2014dha}), whereas in GR these potentials are the same at late time. This property remains of course valid in the presence of a dark fifth force acting on dark matter (only very exotic models of dark matter with a non-zero anisotropic stress would modify this.) Hence measuring the relation between $\Phi$ and $\Psi$, the so-called anisotropic stress, is often refereed to as a smoking gun for modified gravity, i.e.\ a way of distinguishing a dark fifth force from a deviation from GR~\citep{EuclidTheoryWorkingGroup:2012gxx,Song:2010rm,Motta:2013cwa,Amendola:2016saw}.

As discussed in~\cite{Bonvin:2022tii}, this test can however not been performed with current methods. The reason for this is that, besides RSD, we are currently only able to measure the sum of the two gravitational potentials, $\Phi+\Psi$, through gravitational lensing (cosmic shear or CMB lensing)~\citep{Kohlinger:2017sxk,DES:2021wwk}. We have no individual measurements of $\Phi$ or $\Psi$. The only way to measure the anisotropic stress is therefore to use measurements of the galaxy velocity $V$ from RSD, and translate these into measurements of $\Psi$ assuming the validity of Euler equation~\citep{Motta:2013cwa}. While this procedure is perfectly valid in models of modified gravity, it clearly does not work in models with a dark fifth force, that break the equivalence principle. Wrongly assuming Euler equation in this case would lead us to measure an effective difference between $\Phi$ and $\Psi$, and to erroneously conclude that gravity is modified even if it is not. 

In this Letter, we design a method to solve this problem by measuring $\Psi$ directly from large-scale structure observations. 
We use the fact that gravitational redshift leaves an impact on the distribution of galaxies~\citep{Yoo:2009au,Bonvin:2011bg,Challinor:2011bk}, which can be measured by looking for a dipole in the cross-correlation of bright and faint galaxies~\citep{Yoo:2012se,Croft:2013taa,Bonvin:2013ogt,Bonvin:2014owa,Gaztanaga:2015jrs,Bonvin:2015kuc,Beutler:2020evf,Saga:2021jrh}. 
Gravitational redshift is an effect predicted by \cite{Einstein:1916vd}, which postulates that time passes at a slower rate in a gravitational potential well. This effect is directly sensitive to the temporal gravitational potential $\Psi$. It has been measured in a variety of systems: around earth, by comparing clocks at different distances from the ground, i.e.\ at different values of $\Psi$~\citep{PhysRev.140.B788,PhysRevLett.45.2081}; in white dwarf, by looking at the reddening of light due to its escape from the gravitational field of the star~\citep{1925Obs....48..337A,1967AJ.....72Q.301G}; and in clusters of galaxies and non-linear structures by looking at the difference in redshift between galaxies at the centre of the structure and galaxies away from the centre~\citep{Wojtak:2011ia,Sadeh:2014rya,Alam:2017izi}. So far, all these measurements have confirmed the validity of GR. However, this effect (and consequently the potential $\Psi$) has never been measured at cosmological scales. The goal of this paper is to forecast the precision with which $\Psi$ will be measured at cosmological scales with the coming generation of surveys.   

We build on a method presented in~\cite{Sobral-Blanco:2021cks}, where we showed that by combining different multipoles of the galaxy power spectrum, one can isolate the correlation of $\Psi$ with the density. For our forecasts we work with the correlation function, instead of the power spectrum, since the correlation function can account for wide-angle effects, that contaminate the measurement of $\Psi$ and need to be properly removed in order not to bias the measurement. We show that a survey like SKA will be able to measure the evolution of $\Psi$ with a precision of 10--30 percent.

\section{Evolution of the gravitational potential} 
\label{sec:evol}

We start by parameterizing the evolution of the gravitational potential in a model-independent way. We assume that at early time, well in the matter era, the Universe is well described by GR and Cold Dark Matter (CDM). This has been tested with great precision by CMB measurements. We link therefore deviations from the CDM model to the onset of the accelerated expansion of the Universe. In this context, the matter density perturbations in Fourier space are usually modelled as
\begin{align}
\delta(\bk, z)=T_\delta(k,z)\Psi_{\rm in}(\bk)\, ,   
\end{align}
where $\Psi_{\rm in}$ is the primordial gravitational potential generated by inflation and $T_\delta(k,z)$ is the density transfer function, which depends on the Fourier mode $k$ and on the redshift $z$. Under the assumption that deviations from GR appear only at late time, we can write this transfer function as\begin{align}
T_\delta(k,z)=\frac{D_1(k,z)}{D_1(z_*)}T_\delta(k,z_*)\, ,
\end{align}
where $T_\delta(k,z_*)$ is the transfer function predicted by GR at redshift $z_*$, well in the matter era. The growth function $D_1(k,z)$ describes the growth of structure at late time and depends consequently on the theory of gravity and on the model of dark energy. At $z_*$, where GR is recovered, this function is independent on $k$.  

The velocity transfer function defined through $V(\bk, z)=T_V(k,z)\Psi_{\rm in}(\bk)$ is usually related to the density transfer function using the continuity equation for matter:
\begin{align}
T_V(k,z)=-\frac{\HH}{k}f(k,z)T_\delta(k,z)\, ,   
\end{align}
where $\HH$ denotes the Hubble parameter in conformal time and the growth rate $f$ is defined as
\begin{align}
f(k,z)\equiv \frac{d\ln D_1(k,a)}{d\ln a}  \, .   
\end{align}
Under this assumption, the velocity transfer function evolves at late time as
\begin{align}
T_V(k,z)=\frac{\HH(z)f(k,z)D_1(k,z)}{\HH(z_*)f(z_*)D_1(z_*)}T_V(k,z_*)\, .
\end{align}
If the continuity equation is not valid, this standard parameterization breaks down. This would happen for example if dark matter exchanges energy with dark energy~\citep{Pourtsidou:2013nha}. Since we want here to be agnostic about the theory of gravity and the behaviour of dark matter, we parameterize instead the growth of velocity at late time with a new function $G(k,z)$ such that
\begin{align}
T_V(k,z)=\frac{\HH(z)G(k,z)}{\HH(z_*)f(z_*)D_1(z_*)}T_V(k,z_*)\, .
\end{align} 

Finally, the potential transfer function defined through $\Psi(\bk, z)=T_\Psi(k,z)\Psi_{\rm in}(\bk)$ can be related to the density transfer function, using the Poisson equation and assuming that there is no anisotropic stress via
\begin{align}
T_\Psi(k,z)=-\frac{3}{2}\Omega_m(z)\left(\frac{\HH(z)}{k}\right)^2T_\delta(k,z)\, .  
\end{align}
Under these assumptions, the gravitational potential at late time evolves as
\begin{align}
T_\Psi(k,z)=\frac{\HH^2(z)\Omega_m(z)D_1(k,z)}{\HH^2(z_*)D_1(z_*)}T_\Psi(k,z_*)\, ,
\end{align}
where we have used that $\Omega_m(z_*)=1$.
Again, in order to be agnostic about the theory of gravity, we replace this standard evolution with a new free function $\I(k,z)$ which encodes the evolution of the gravitational potential $\Psi$ at late time:
\begin{align}
T_\Psi(k,z)=\frac{\HH^2(z)\I(k,z)}{\HH^2(z_*)D_1(z_*)}T_\Psi(k,z_*)\, .
\end{align}

The functions $D_1(k,z)$ and $G(k,z)$ have been measured via redshift-space distortions. The function $I(k,z)$ on the other hand has never been measured at cosmological scales. The goal of this paper is to forecast the precision with which one can measure this function with the coming generation of surveys.

\section{Correlation function} 
\label{sec:corr}

Redshift surveys map the distribution of galaxies in the sky and measure the galaxy number counts fluctuations
\be
\Delta\equiv\frac{N(\bn,z)-\bar N(z)}{\bar N(z)}\, ,
\ee
where $N$ is the number of galaxies per pixel detected in direction $\bn$ and at redshift $z$, and $\bar N$ denotes the average number of galaxies per pixel at redshift $z$. In the linear regime, the dominant contributions to the observable $\Delta$ are given by~\citep{Bonvin:2011bg,Challinor:2011bk,Yoo:2009au}
\begin{align}
\Delta(\bn, z)&=b\,\delta-\frac{1}{\HH}\partial_r(\bV\cdot\bn)+\frac{1}{\HH}\partial_r\Psi+\frac{1}{\HH}\dot\bV\cdot\bn \label{Delta}\\
&+\left(1-5s+\frac{5s-2}{r\HH}-\frac{\dot\HH}{\HH^2}+f^{\rm evol}\right)\bV\cdot\bn\, ,\nonumber
\end{align}
where $r=r(z)$ is the comoving distance to redshift $z$ and a dot denotes derivative with respect to conformal time. The  functions $b(z)$, $s(z)$  and $f^{\rm evol}(z)$  are the  galaxy bias, the magnification bias and the evolution bias of the population under consideration.

The two terms in the first line of Eq.~\eqref{Delta} are the standard density and redshift-space distortions contributions. The third term is the contribution from gravitational redshift, which depends directly on $\Psi$ and is the target of our study. The last two terms are Doppler contributions~\footnote{Note that other relativistic distortions contribute to $\Delta$ \citep{Bonvin:2011bg, Challinor:2011bk, Yoo:2009au}, but their impact on the multipoles of the correlation function is negligible~\citep{Jelic-Cizmek:2020pkh}. Gravitational lensing also contributes to $\Delta$, but its impact becomes important only at large redshifts~\citep{Jelic-Cizmek:2020pkh,Euclid:2021rez}, where we will see that $I$ cannot be measured anymore.}.

Density and RSD generate a monopole, quadrupole and hexadecapole in the correlation function, which have been measured with various surveys, see e.g.~\cite{eBOSS:2020yzd}. The contributions from gravitational redshift and Doppler effects on these multipoles are strongly subdominant and can safely be neglected~\citep{Jelic-Cizmek:2020pkh}. However, these relativistic effects generate a dipole in the correlation function, which can be measured by correlating two populations of galaxies, for example bright and faint galaxies~\citep{Bonvin:2013ogt}. We now write the multipoles in terms of the free functions $D_1, G$ and $I$. The even multipoles take the form
\begin{align}
\xi_0(z,d)&=\left[\hat{b}^2(z) +\frac{2}{3}\hat b(z)\hat G(z)+\frac{1}{5}\hat G^2(z)\right]\mu_0(z_*, d)\, ,\nonumber\\  \xi_2(z,d)&=-\left[\frac{4}{3}\hat{b}(z)\hat{G}(z)+\frac{4}{7}\hat{G}^2(z)\right]\,\mu_2(z_*,d)\, ,\nonumber\\
\xi_4(z,d)&= \frac{8}{35}\hat{G}^2(z)\,\mu_4(z_*,d)\, ,\label{eq:multeven}
\end{align}
where
\begin{align}
\hat b(z)&\equiv b(z)\sigma_8(z)\,  ,\\
\hat G(z)&\equiv G(z)\frac{\sigma_8(z)}{D_1(z)}=G(z)\frac{\sigma_8(z_*)}{D_1(z_*)}\, ,
\end{align}
and
\begin{align}
\mu_\ell(z_*,d)&=\frac{1}{2\pi^2}\int dk k^2 \frac{P_{\delta\delta}(k,z_*)}{\sigma_8^2(z_*)}j_\ell(kd)\, . \label{eq:muell}
\end{align}
The functions $\mu_\ell(z_*,d)$ depend on the matter power spectrum at $z=z_*$, well in the matter era, and are therefore determined by physics in the early Universe, which is tightly constrained by CMB observations~\citep{Planck:2018vyg}. These functions are therefore considered as fixed. Measuring the even multipoles provides consequently direct measurements of the bias, $\hat b(z)$, and of the function $\hat G(z)$~\citep{eBOSS:2020yzd} (which reduces to $f(z)\sigma_8(z)$ under the assumption that the continuity equation is valid.) 

The dipole in the cross-correlation of bright and faint galaxies reads~\footnote{Note that the dipole is also affected by wide-angle effects, which generate a contribution of the form $-2/5(\hat{b}_\B-\hat{b}_\F)\hat{G}d/r\mu_2(z_*,d)$. This contribution can however be removed by constructing an appropriate estimator, as has been shown in~\cite{Bonvin:2013ogt,Hall:2016bmm,Bonvin:2018ckp}. For simplicity we have set here the evolution biases to zero $f^{\rm evol}_\B=f^{\rm evol}_\F$. In practice these parameters will be measured from the galaxy populations, see e.g.~\cite{Wang:2020ibf}.}
\begin{align}
\xi_1(z,d)&= \frac{\HH}{\HH_0}\Bigg\{3\left(\frac{1}{r\HH}-1 \right)\big(s_\B(z)-s_\F(z)\big)\hat{G}^2(z) \nonumber\\
&\quad+5\left(\frac{1}{r\HH}-1 \right)\left(\hat{b}_\F(z)s_\B(z)-\hat{b}_\B(z)s_\F(z)\right)\hat{G}(z) \nonumber\\
&\quad+\left(\hat{b}_\B(z)-\hat{b}_\F(z)\right)\left[\left(\frac{2}{r\HH}-1\right)\hat{G}(z)-\frac{\Dot{\hat{G}}(z)}{\HH}\right]\nonumber\\
&\quad+\frac{3}{2}\left(\hat{b}_\B(z)-\hat{b}_\F(z)\right)\hat{I}(z)\Bigg\}\,\nu_1(z_*,d)\, , \label{eq:dip}
\end{align}
where
\begin{align}
\hat I(z)&\equiv I(z)\frac{\sigma_8(z)}{D_1(z)}=I(z)\frac{\sigma_8(z_*)}{D_1(z_*)}\, ,
\end{align}
and
\begin{align}
\nu_1(z_*,d)&=\frac{\HH_0}{2\pi^2}\int dk k \frac{P_{\delta\delta}(k,z_*)}{\sigma_8^2(z_*)}j_1(kd)\, . \label{eq:nu1}
\end{align}
As before, the function $\nu_1(z_*,d)$ which depends on $z_*$ is considered fixed. Combining the dipole with the even multipoles provides a way of measuring directly the function $\hat I(z)$, which encodes the evolution of the gravitational potential $\Psi$.

Note that in our derivation we have assumed that the $k$-dependence of $D_1, G$ and $I$ is negligible, so that we can take them out of the integrals in Eqs.~\eqref{eq:muell} and~\eqref{eq:nu1}. This is a common assumption, that is used in many analyses, see e.g.~\cite{eBOSS:2020yzd}, and that is motivated by the fact that in the quasi-static approximation the $k$-dependence can often be neglected~\citep{Gleyzes:2015rua,Raveri:2021dbu}. However, for some theories of gravity this assumption may not be valid~\citep{Baker:2014zva}. Relaxing it would slightly complicate the forecasts, but would not invalidate our method. 

\section{Forecasts} 
\label{sec:forecasts}

We now forecast how well we can measure $\hat I$ with a survey like SKA phase 2, using the Fisher formalism. SKA2 will observe close to a billion galaxies from $z=0.1$ to $z=2$. We use the specifications of~\cite{Bull:2015lja}, for the number density and volume, and the cosmology is fixed to the latest \cite{Planck:2018vyg} values. 
We choose $z_*=10$, well in the matter era, and as described in Section~\ref{sec:corr} we assumed that the functions $\mu_\ell(z_*,d)$ and $\nu_1(z_*,d)$ are fixed by CMB constraints. Therefore we vary only $\hat{G}$, $\hat{I}$ and the biases of the bright and faint populations in each redshift bin~\footnote{Note that the derivatives in the Fisher matrix can be computed analytically for our set of parameters.}. We split the galaxies into a bright and faint population with same number of galaxies per redshift bin. 
We model the bias for each of the population as
\begin{align}
b_{\B}=c_{\B} \exp(d_{\B}z)+\Delta b/2\quad\mbox{and}\quad  b_{\F}=c_{\F} \exp(d_{\F}z)-\Delta b/2\label{eq:bias}
\end{align}
involving four parameters with values $c_\B=c_\F = 0.554$ and $d_\B=d_\F = 0.783$~\citep{Bull:2015lja}, and a bias difference $\Delta b=1$, similar to what has been measured for BOSS in~\cite{Gaztanaga:2015jrs}. We then let the biases $\hat{b}_\B=b_\B \sigma_8$ and $\hat{b}_\F=b_\F\sigma_8$ vary freely in each redshift bins and marginalise over them.
For the magnification bias, we use the model developed in~\cite{Castello:2022uuu}, and we assume that the evolution bias vanishes for the two populations. 
Once data will be available the magnification bias and the evolution bias of the two populations will be directly measurable, see e.g.~\cite{Wang:2020ibf}. 
We include shot noise and cosmic variance in the variance of the multipoles (see Appendix~C of~\cite{Bonvin:2018ckp}) and account for cross-correlations between different multipoles.

We perform two different forecasts: the first one where we consider as free parameters the functions $\hat G$ and $\hat I$ in each redshift bin, and a second one where we parameterize the evolution of these functions with redshift. 

\subsection{Constraints per redshift bin}

\begin{table}
\caption{$1\sigma$ constraints on $\hat I$ and $\hat G$ relative to their corresponding fiducial value, marginalised over the bias parameters. We show the results for two values of the minimal separation $d_{\rm min}$, in Mpc/$h$.}
\label{table:1}
\centering
\begin{tabular}{C{0.4cm} C{0.2cm} C{0.45cm} C{0.45cm} C{0.45cm} C{0.45cm} C{0.45cm} C{0.45cm} C{0.45cm} C{0.45cm}} 
 \hline
 &$z$ & 0.35 & 0.45 & 0.55 & 0.65 & 0.75 & 0.85 & 0.95 & 1.05 \\
 \hline
 $d_{\rm min}$&$\hat{I}$\Tstrut & 0.23 & 0.24 & 0.28 & 0.33 & 0.39 & 0.48 & 0.60 & 0.77 \\ 
 20 &$\hat{G}$ & 0.002 & 0.002 & 0.003 & 0.003 & 0.003 & 0.004 & 0.004 & 0.005 \\
 \hline
 $d_{\rm min}$&$\hat{I}$\Tstrut & 0.27 & 0.28 & 0.32 & 0.37 & 0.45 & 0.55 & 0.69 & 0.87 \\ 
32 &$\hat{G}$ & 0.004 & 0.004 & 0.005 & 0.006 & 0.006 & 0.007 & 0.008 & 0.009 \\  
 \hline
\end{tabular}
\end{table}

From Eq.~\eqref{eq:dip}, we see that the dipole depends not only on $\hat G$ in each redshift bin, but also on the time derivative of $\hat G$. 
Since we do not want to assume any evolution for $\hat G$, we express this time derivative in terms of $\hat G$ in the neighbouring bins. We use the five-point stencil method:
\begin{align}
&\dot{\hat{G}}(z_i)=-(1+z_i)\HH(z_i)\frac{d\hat G(z_i)}{dz_i}=-\frac{(1+z_i)\HH(z_i)}{12\Delta z}\times \nonumber\\
&\Bigg[-G(z_{i+2})+8G(z_{i+1})-8G(z_{i-1})+G(z_{i-2})\Bigg] \, . 
\end{align}
We have checked that for redshift bins of size $\Delta z=0.1$, the five-point stencil method allows us to reconstruct $\dot{\hat{G}}(z_i)$ with a precision of $0.1$ percent. 
On the other hand, if we use only two bins to reconstruct the time derivative we would make an error of up to 129 percent at low redshift, due to the fact that $\dot{\hat{G}}$ changes sign. 
Such a large mistake in $\dot{\hat{G}}$ may bias the measurement of $\hat I$.

In Table \ref{table:1}, we show the 1$\sigma$ constraints on $\hat I$ and $\hat G$, marginalised over the bias parameters. 
Note that due to the five-point stencil method we cannot constrain $\hat I$ in the first two and in the last two redshift bins, since in these four bins $\dot{\hat{G}}$ is not constrained. 
We show therefore the constraints starting at $z=0.35$. We compute the constraints starting at two different minimum separations: $d_{\rm min}=20\,\mathrm{Mpc}/h$ and $d_{\rm min}=32\,\mathrm{Mpc}/h$, since non-linearities have been shown to become relevant around those scales~\citep{Bonvin:2020cxp}. Since the signal-to-noise ratio peaks around 30\,Mpc$/h$ and then slowly decreases with separation (see e.g.\ Fig. 5 in~\cite{Bonvin:2018ckp}), separations above $d_{\rm max}=160\,\mathrm{Mpc}/h$ are irrelevant. The constraints are therefore obtained from separations well inside the horizon, which justifies the use of the quasi-static approximation.

\begin{figure}
    \centering
    \includegraphics[scale=0.47]{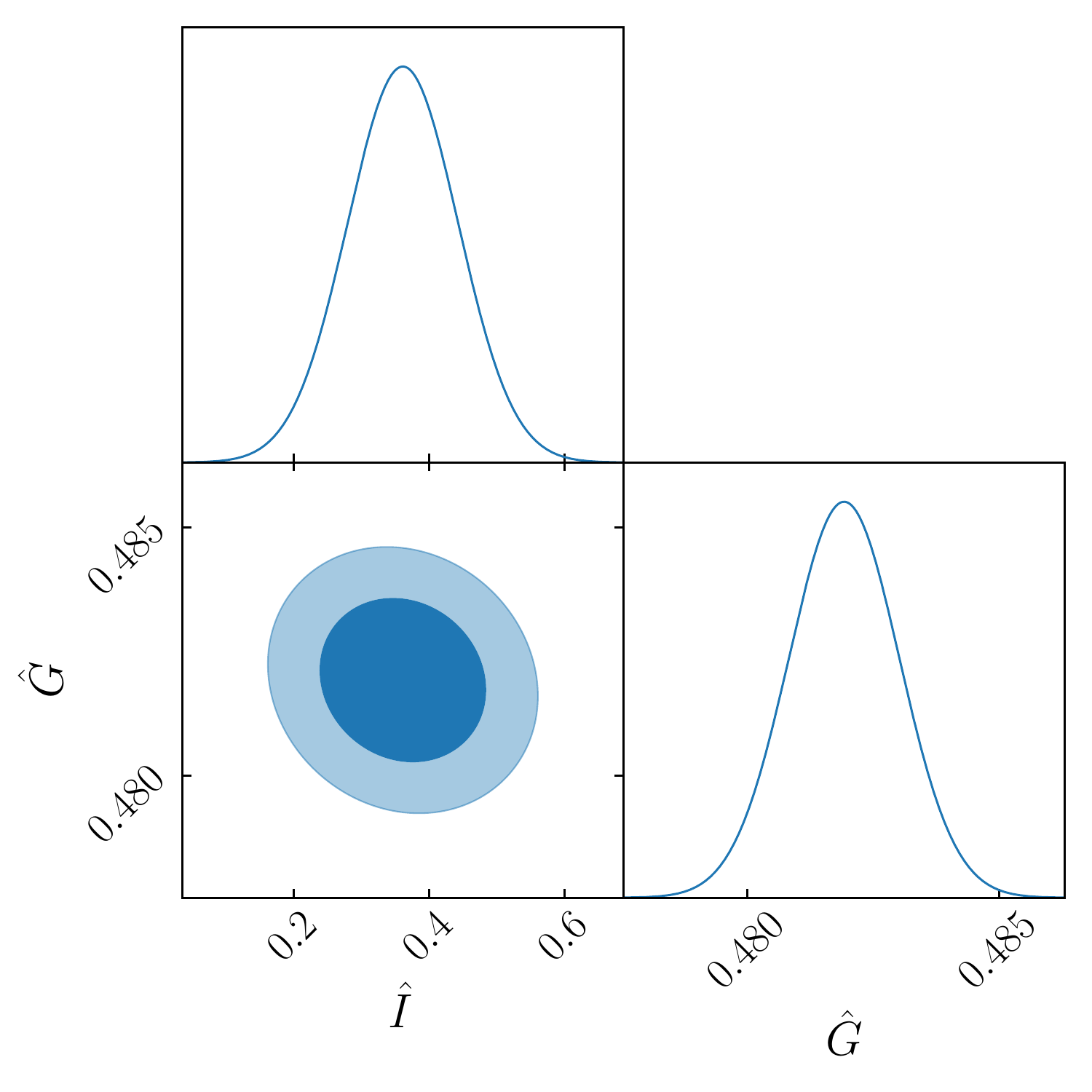}
    \caption{Joint constraints of $\hat{I}$ and $\hat{G}$ at $z=0.35$, with $d_{\rm min}=20\,\mathrm{Mpc/}h$.}
    \label{fig:plotIandGz3}
\end{figure}

We see that at low redshift, the constraints on $\hat I$ are very good, providing a direct measurement on the evolution of the potential $\Psi$ with a precision of 20-30 percent. Combining such a measurement with gravitational lensing will allow us to distinguish unambiguously between deviations from GR and dark fifth force models. Comparing with current tests of gravity, we see that the precision with which $\Psi$ will be measured with the coming generation of survey is similar to the precision with which the evolution of the velocity field is currently measured: in~\cite{eBOSS:2020yzd}, for example, the combination $f(z)\sigma_8(z)$ is measured with a precision of 10-30 percent in various redshift bins. The future constraints on $\hat G$ are tighter than those on $\hat I$ by almost two order of magnitude. 
This is not surprising since $\hat G$ is constrained by the even multipoles, which will be measured with exquisite accuracy with SKA2, whereas $\hat I$ is constrained only by the dipole, which has a significantly lower signal-to-noise ratio. Interestingly, the constraints on $\hat I$ are only mildly degraded (by $14-18\%$) when increasing $d_{\rm min}$ from 20 to 32\,Mpc/$h$. In contrast, the constraints on $\hat G$ are degraded by 100\%. This is due to the fact that the gravitational potential is much less affected by non-linearities than density and RSD.

In Fig.~\ref{fig:plotIandGz3} we show the joint constraints on $\hat G$ and $\hat I$ at $z=0.35$. 
We see that the parameters are almost not degenerate. This is due to the fact that the even multipoles are only sensitive to $\hat G$ and can therefore efficiently break the degeneracy between $\hat G$ and $\hat I$ in the dipole.

Finally, we have also performed forecasts for the parameter $\tilde I\equiv(\hat{b}_\B-\hat{b}_\F)\hat I$ which directly enters into the dipole. The constraints on this parameter are only very slightly better than those on $\hat I$ (by 0.1 percent). Hence even though $\hat I$ is completely degenerate with the biases in the dipole, these biases are so well constrained by the even multipoles, that the degeneracy is completely broken. 

\subsection{Constraints from a specific time evolution}

When constraining deviations from GR, standard RSD analyses usually assume a specific time evolution for the functions encoding these deviations. One common assumption is to use that the deviations evolve proportionally to the amount of dark energy $\Omega_\Lambda(z)$, see e.g~\cite{Alam:2020jdv}. 
Here we consider the two following models for the evolution of $\hat G$ and $\hat I$:
\begin{align}
\hat G(z)=&f(z)\sigma_8(z)\left[1+ \hat{G}_0X(z)\right]\, ,  \label{eq:evolG}  \\
\hat I(z)=&\Omega_m(z)\sigma_8(z)\left[1+ \hat{I}_0X(x)\right]\, , \label{eq:evolI}  
\end{align}
with $X(z)=1$ for $z\in [0,2]$ in the first model and $X(z)=\Omega_\Lambda(z)/\Omega_\Lambda(z=0)$ in the second model. 
With this, the analysis is significantly simplified since $\dot{\hat G}$ is directly determined by Eq.~\eqref{eq:evolG}. Moreover in this case, instead of considering one free bias parameter per redshift for each population, we assume that the biases evolve as in Eq.~\eqref{eq:bias}, with four free parameters $c_\B, c_\F, d_\B$ and $d_\F$.

The marginalised constraints on $\hat G_0$ and $\hat I_0$ are shown in Table~\ref{table:2}. For the constant model, we reach a precision of 10\% on $\hat I_0$, which decreases to 15\% for the dynamical model. 

\begin{table}
\caption{$1\sigma$ constraints on $\hat{I}_0$ and $\hat{G}_0$, marginalised over the bias parameters, for the two models described after Eq.~\eqref{eq:evolI}.}
\label{table:2}
\centering
\begin{tabular}{@{}c c c c c@{}}
  \hline
  & \multicolumn{2}{c}{No evolution}&\multicolumn{2}{c}{Evolution}\\
  $d_{\rm min} [{\rm Mpc}/h]$\Bstrut& 20 & 32 & 20 & 32 \\ 
 \hline
 $\hat{I}_0$\Tstrut & 0.10 & 0.11 & 0.15 & 0.18  \\
 $\hat{G}_0$ & 0.001 & 0.002 & 0.002 & 0.003 \\
 \hline
\end{tabular}
\end{table}

\section{Conclusion} 
\label{sec:conclusion}

In this Letter we have shown that the coming generation of galaxy surveys, like SKA2, will be able to measure directly the evolution of the gravitational potential, $\Psi$, with an accuracy of 10--30\%. Such a measurement will be a game changer to distinguish modified theories of gravity from models with a dark fifth force, since it is fully complementary to current measurements that are sensitive to the matter density and velocity, and to the sum of the two gravitational potentials. 

One difficulty in measuring the evolution of $\Psi$ is that it is degenerate with the time derivative of the velocity, which is itself unknown. However, we have shown that since future surveys will be able to measure the velocity in thin redshift bins, this can be used to reconstruct, in a model-independent way, the time derivative of the velocity and consequently break the degeneracy with $\Psi$.

Comparing our forecasts with current constraints on modified gravity, we see that future constraints on $\Psi$ are roughly of the same order of magnitude as current constraints on the velocity $V$, and worse by two orders of magnitude than future constraints on $V$. This is simply due to the fact that the dipole (which is constraining $\Psi$) has a signal-to-noise ratio which is significantly lower than that of the even multipoles (which are constraining $V$). However, it is worth mentioning that the very tight constraints expected on $V$ do not directly translate into tight constraints on modified gravity parameters, like for example $\mu_0$ (which encodes a modification to Poisson equation~\citep{Pogosian:2010tj}). 
As shown in~\cite{Castello:2022uuu}, if one does not assume that dark matter obeys the weak equivalence principle, $\mu_0$ can only be constrained with a precision of 15\% with a survey like SKA2, which is very similar to the constraints we found here on $\Psi$. This seems therefore to be the level at which modifications to GR can be tested in a truly model-independent way.

A first application of this novel measurement of $\Psi$, will be to use it in combination with gravitational lensing to measure the anisotropic stress in a model-independent way.

\section*{Acknowledgements}
 This project has received funding from the Swiss National Science Foundation and from the European Research Council (ERC) under the European Union’s Horizon 2020 research and innovation program (Grant agreement No.~863929; project title ``Testing the law of gravity with novel large-scale structure observables"). We acknowledge the use of the fftlog-python code written by Goran Jelic-Cizmek and available at \url{https://github.com/JCGoran/fftlog-python}.

\section*{Data Availability}

No new data were generated in support of this research. The numerical results underlying this article will be shared on
reasonable request to the corresponding author.



\bibliographystyle{mnras}
\bibliography{isolate_psi} 








\bsp	
\label{lastpage}
\end{document}